\newcommand{\be}{\begin{eqnarray}}
\newcommand{\ee}{\end{eqnarray}}

\def\MeV{{\rm MeV}}
\def\GeV{{\rm GeV}}

\def\lsim{\mathrel{\rlap{\lower3pt\hbox{\hskip1pt$\sim$}}
     \raise1pt\hbox{$<$}}} 
\def\gsim{\mathrel{\rlap{\lower3pt\hbox{\hskip1pt$\sim$}}
     \raise1pt\hbox{$>$}}} 

\def\barc{$\bar c c\,\,\,$}
\def\barQ{$\bar Q Q$ }
\def\Jp{$J/\psi$ }

\pdfoutput=1
\documentclass[twocolumn,letter,showpacs,preprintnumbers,amsmath,amssymb,floatfix,nofootinbib]{revtex4}

\usepackage{graphicx}
\usepackage{dcolumn}
\usepackage{bm}

\begin{document}


\title{Recombinant Charmonium in strongly coupled
  Quark-Gluon Plasma}

\author{  Clint Young and Edward Shuryak}
\address { Department of Physics and Astronomy\\
State University of New York, Stony Brook, NY 11794-3800 }

\date{\today}
\begin{abstract}
We update our previous work of a Langevin-with-interaction model for 
charmonium in heavy-ion collisions, by considering the effect due to recombination.
We determine the contribution to \Jp yields from \barc  pairs whose constituent quarks 
 originate from two different hard processes. Like the surviving \Jp states, the recombinant \Jp
 also undergo both a stochastic interaction, determined 
by a hydrodynamical simulation of the heavy-ion collision, and an interaction determined 
by the \barQ potentials measured on the lattice for appropriate temperatures. From the results of these simulations, we determine 
both the direct and the recombinant contribution to the \Jp yields for RHIC conditions, and find that 
for central collisions, between 30\% and 50\% of the \Jp yield is due to recombinant production. 
We compare our results with other models and look for how the recombinant contribution 
differs from the surviving contribution in the differential $p_T$ yields.
Including the recombinant contribution improves the agreement 
with  the latest analysis of charmonium at RHIC, which shows an absence of anomalous 
suppression except in the most central collisions.

\end{abstract}
\maketitle
\begin{narrowtext}

\section{Introduction}

In a previous paper \cite{Young:2008he}, we argued that the microscopic dynamics of 
charmonium in a heavy-ion collision should be modeled as a stochastic interaction with 
strongly-coupled quark-gluon plasma (sQGP). In such a plasma the diffusion coefficient is 
very small, leading to rapid thermalization in momentum space and slow spatial diffusion, 
which is further slowed by the attraction of the constituent charm and anti-charm 
quarks in the pair. We concluded that the amount of equilibration in space
(and thus the \Jp yield) depends strongly 
on the timescales of the collision.  In realistic simulations of Au+Au collisions at 
RHIC, where the sQGP phase exists for $\tau \sim 5\; {\rm fm}/c$ , our model 
predicted a survival probability for \Jp $\sim 1/2$ even in the most central collisions, much larger than previously expected. Those results
were able to explain qualitatively the data from PHENIX.
 
In \cite{Young:2008he}, we did not consider another  possibly important source of charmonium at RHIC, the ``recombinant'' contribution of \Jp particles whose constituent quarks originate from different 
hard processes. At very high collision energies, as the number of charm pairs per event grows, 
recombinant charmonia could potentially lead to an 
enhancement of the final \Jp yields in a heavy-ion collision, reversing the current
suppression trend. Using the grand canonical ensemble 
approach, Braun-Munzinger and collaborators \cite{BraunMunzinger:2000px} determine
the fugacity of charm  by the number of \barc pairs produced initially. The ``statistical hadronization" 
approach to charmonium assumes complete thermal equilibration of charmonium.  
 Another approach has been taken by Grandchamp and Rapp
\cite{Grandchamp:2002wp}, 
who treat the \Jp yields from heavy-ion collisions as coming from two sources:  the direct 
component, which decays exponentially with some lifetime $\tau_d$, and  the coalescent  
component, which is determined by the same mechanism in \cite{BraunMunzinger:2000px}, with 
the additional consideration that spatial equilibration of charm does not happen. To account for 
enhanced local charm density, by small spatial diffusion, they had introduced another factor - the ``correlation volume" $V_{corr}$ - which was estimated.
The present work can be viewd as a quantitative dynamical calculation of this parameter.

To gain insight, we should compare these models with our model in \cite{Young:2008he}.
The Langevin-with-interaction model for \barc pairs in medium makes no assumptions about 
complete thermalization, and shows how even in central Au+Au collisions at the RHIC, the 
majority of the \Jp yields may survive the QGP phase. However, the 
model  predicts rapid thermalization in the momentum 
distributions of charmonium, as well as  
equilibration in the relative yields of the various charmonium states due to the 
formation of ``quasi-equilibrium" in phase space. This requires no fine-tuning of the rates for charmonium in plasma; it is 
just a natural consequence of the strongly coupled
nature of the media, detailed by the Langevin dynamics. However, recombinant production of charmonium may still be an 
important effect in our model, due to the fact that in central collisions, the densities of unbound 
charm quarks can be quite high in some regions of the transverse plane.

Our model simulates an ensemble of \barc pairs, generated initially by PYTHIA event generation
and then evolved according to the Langevin-with-interaction model.
We evolve the pairs  
not assuming any form of equilibrium, and then average over possible pairings of the quarks to 
form recombinant charmonium.

The outline of this work is as follows: in Section 
\ref{simulation} we will describe how we simulated charm in plasma and took into account the 
contribution due to recombinant \Jp, and in Section \ref{conclusions} we take the opportunity to describe 
the progress in this model so far and also to summarize where future work is needed for 
a Langevin-with-interaction description of \Jp suppression. In Appendix \ref{canonical}, we 
discuss the statistics necessary to calculate the recombinant contribution to \Jp 
yields.

\section{Recombinant charmonium in heavy-ion collisions}

\label{simulation}

\subsection{Langevin-with-interaction model for \barc pairs in a heavy-ion collision}

As we have done in our previous paper, we  simulate \Jp in medium with a hydrodynamical 
simulation of the collision. As before, we start with a large ensemble of \barc pairs whose momenta 
are determined with PYTHIA event generation \cite{Andersson:1977qx}. The  positions 
of the initial hard collisions in the transverse plane at 
mid-rapidity are determined by sampling the distribution in $N_{coll}$ determined from the 
Glauber model. In this way, our local densities of \barc pairs vary as one would expect from the 
Glauber model, which gives an enhancement for recombination towards the center of the transverse 
plane. Each element of the ensemble 
now contains $N$ \barc pairs. The number of pairs $N$ depends on the impact parameter of the 
collision and needs to be determined.

The average number of \barc pairs for a Au+Au collision at RHIC varies with impact parameter and has 
been investigated by the PHENIX collaboration at mid-rapidity \cite{Adler:2004ta}. The measured
 cross sections for charm production vary somewhat with the centrality of the collision and 
 achieves a maximum of 
about $800\; \mu b$ for semi-central collisions. The nuclear overlap function 
$T_{AA}(b)$ can be calculated with the Glauber model. We used a convenient program by Dariusz Miskowiec \cite{miskowiec} to evaluate this function. With a centrality dependent cross-section $\sigma_{\bar{c}c}$, we can easily calculate the 
average number of \barc pairs in a collision: $N_{\bar{c}c}=T_{AA}\sigma_{\bar{c}c}$.
The number of \barc pairs reaches a maximum in central collisions, with an average of 19 pairs per 
collision.

In order to determine the probability for two charm quarks from different hard processes to form 
recombinant charmonium, we must average over the different possible pairings of all of the unbound 
quarks in each element of our ensemble. This is discussed in Appendix \ref{canonical} in 
generality. 
Since the number of \barc pairs approaches 20 for central Au+Au collisions at RHIC, we are faced with 
another issue: there are 
20!  possible pairings  and it has become impractical to 
calculate the probability of each individual pairing this way. In general, we would be forced to perform 
{\it permutation sampling} of this partition function, preferably with some Metropolis algorithm. How to sample over permutations with a Metropolis algorithm is discussed thoroughly in the literature, for an excellent review of this see Ceperley \cite{ceperley}. 
However, for  RHIC, the situation simplifies due to the relatively low densities 
of \barc pairs involved.  We ran our simulation for the most central Au+Au collisions at RHIC and 
examined how many ``neighbors" each charm quark had.  A ``neighbor" is defined as a charm 
anti-quark, originating from a different pQCD event yielding the given charm quark, which is close 
enough to the charm quark that it could potentially form a bound state, in other words $r$ is such that
$V_{cornell}(r)<0.88\;\GeV$. The number 
of charm quarks expected to have one and only one neighbor in the most central Au+Au collisions was found to be  5.5\%,
while only 0.2\% of the charm 
quarks are expected to have more than one neighbor. 
Therefore, even at the most central collisions at RHIC, we can be spared possibly complicated permutation 
samplings. Of course, this situation is not true in general, and for the numbers of pairs produced in a 
typical heavy-ion collision at the LHC one should modify these combinatorics.

\subsection{New analysis of the data including improved $dAu$ sample}

The data with which we now compare our results is different from that which we 
used for comparison in our previous work. New data analysis of Au+Au and d+Au described in
\cite{leitch} account for the (anti-)shadowing and the breakup of 
charmonium due to the cold nuclear matter effects (parameterized by $\sigma_{abs}$) 
in the framework of a Glauber model for the collision. The calculations 
at forward and mid-rapidity are now done independently, since shadowing and breakup could be 
considerably different at different rapidities. This new analysis is a significant success, demonstrating 
the high suppression at forward rapidity (previously very puzzling) as being due to cold 
nuclear matter effects. New ratios of observed suppression due to cold nuclear matter 
$R_{AA}/R_{AA}^{CNM}$, plotted versus 
the energy density times time $\epsilon \tau$,  show common trends for RHIC and SPS
 data, which was not the case previously. We use this new analysis as a measure of survival probability in our calculation.

\begin{figure}
\includegraphics[width=8cm]{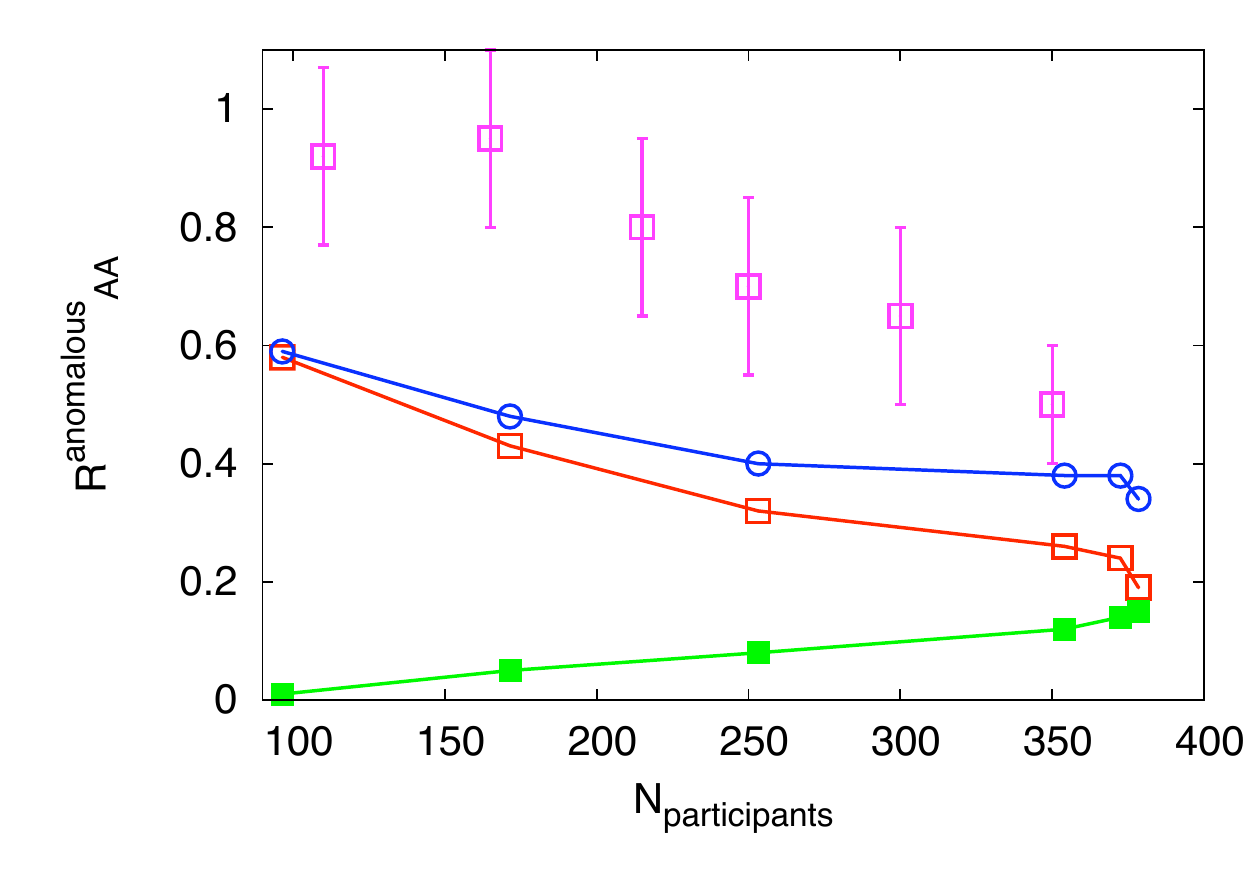}
\includegraphics[width=8cm]{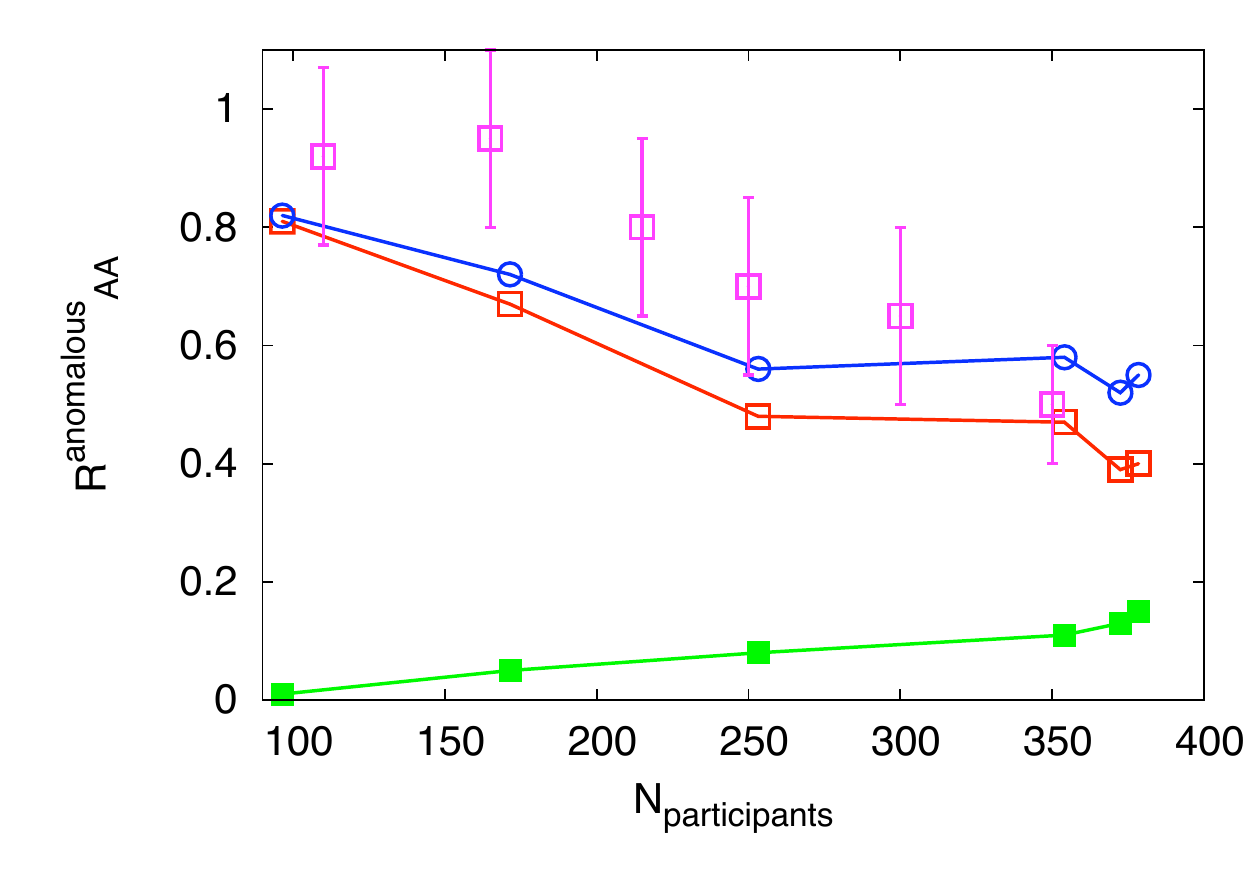}
\caption{ (Color online.)
$R^{anomalous}_{AA}=R_{AA}/R_{AA}^{CNM}$ for \Jp 
versus centrality of the AuAu collisions at RHIC. The data points with error bars show the PHENIX Au+Au measurements 
with cold nuclear matter effects factored out as in \cite{leitch}.
Other points, connected by lines, are our calculations for the two values of the
QCD phase transition temperature $T_c=165\; \MeV$ (upper) and  $T_c=190\; \MeV$ (lower).
 From bottom to top: the (green) filled squares 
show our new results, the  recombinant \Jp, the open (red) squares show the $R_{AA}$ for surviving diagonal 
\Jp, the open (blue) circles show the total.   } 
\label{R_AA_rec}
\end{figure}

\subsection{The results}

Before we show the results, let us remind the reader that our calculation is intended to be a dynamical one,
with no  free parameters. We use a hydrodynamical simulation developed in 
\cite{Teaney:2001av} which 
is known to describe accurately the radial and elliptic collective flows observed in heavy-ion collisions. Our
drag and random force terms for the Langevin dynamics has one input -- the diffusion coefficient for 
charm --  constrained by two independent measurements ($p_T$ distributions and $v_2(p_T)$ measurements for single
lepton -- charm -- performed in Ref.
\cite{Moore:2004tg}. The interaction of these charm quarks are determined by the correlators for two Polyakov lines 
in lQCD \cite{bielefeld}.  

Having said that, we still are aware of certain uncertainties in all the input parameters, which affect the results.
In order to show by how much the results change if we vary some of them, we have used 
 the uncertainty in the value for the critical temperature $T_c$. 
For these reasons, we show  the results for 
two values
$T_c=165,\;190\;\MeV$, in Fig.\ref{R_AA_rec}.

As can be seen, a higher $T_c$ value improves the agreement of our 
simulation with the latest analysis of the data, because in this case
the QGP phase is shorter in duration and the survival probablity is larger.
 However the recombinant contribution
(shown by filled squares) is in this case relatively smaller, making less than 1/3 of the yield even in the most central collisions at RHIC.

Our results for the total, direct, and recombinant contributions resembles considerably the results 
of Zhao and Rapp obtained from their two-component model \cite{Zhao:2008pp}. However it is 
important to point out two important differences. First of all, what is described by Zhao and Rapp as the second component due 
to statistical coalescence includes, with the recombinant \Jp, surviving \Jp, 
destroyed by the medium, which ultimately coalesce in the end. Second, the direct \Jp states' 
abundance, when compared with the abundances of excited charmonium states, does not necessarily 
need to be as expected from these particles' Boltzmann factors. For our model, these relative 
abundances do make sense for direct charmonium states, due to the formation of a quasi-equilibrium 
distribution.

\subsection{Recombinant \Jp and $p_t$-distributions}
\label{Jp_pt}

So far, we have only considered the effect of the recombinant production on the overall yields of 
\Jp particles at the RHIC. We should test our model by considering whether or not adding the 
recombinant contribution can change the shape of any differential \Jp yields.

One differential yield where we may expect the surviving and recombinant component to have different 
behaviors is in the $p_T$-distributions for central Au+Au collisions. The surviving \Jp states tend to 
originate in the periphery of the collision region, since the \Jp states produced here endure the 
sQGP phase for the shortest times. However, the recombinant contribution should form toward the 
center of the collision region, since this is where the density of initial \barc pairs is highest, and as 
we have been showing for some time, spatial diffusion is incomplete in the sQGP. Therefore, since 
the effect due to flow on the $p_T$-distributions has Hubble-like behavior, with the radial velocity 
of the medium scaling with distance from the center of the transverse plane, we would expect 
the recombinant contribution to exist, on average, in regions of the medium with significantly 
smaller flow velocities.

\begin{figure}
\includegraphics[width=8cm]{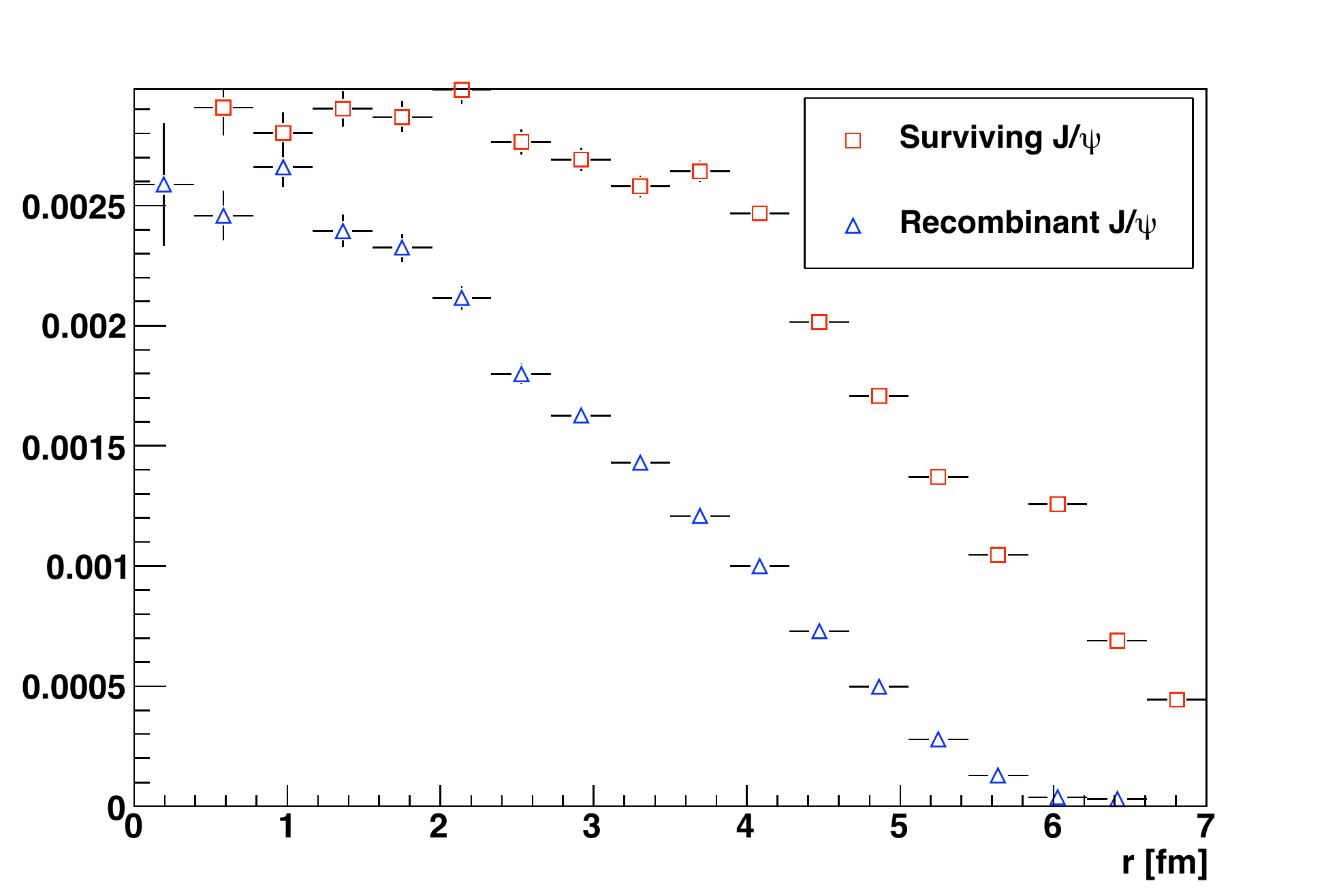}
\caption{ (Color online.)
The surviving and recombinant \Jp yields, plotted versus the radial distance from the center of 
the transverse plane.
 }
\label{r_compare}
\end{figure}

Figure \ref{r_compare} demonstrates this behavior existing in our simulation.

We should now determine whether or not this difference of the yield versus $r$ can be observed in the \Jp yield versus $p_t$. As we have shown in 
our previous paper, during the phase transition from QGP to the hadronic phase in heavy-ion collisions, 
our model predicts a small change in the total \Jp yield but relatively large changes in the \Jp $p_t$
distributions, with these changes strongly dependent on the drag coefficient for quarkonium during 
this time, and $T_c$. We can easily run our code with an LH8 equation of state 
and make several predictions for the two components' $p_t$ distributions. However, for reasons which 
will become apparent,  we are only interested in the upper limit of the 
effect of flow on $p_t$ in Au+Au collisions at the RHIC. Therefore, we ran our simulation where we assumed a phase transition which 
lasts $5 \; {\rm fm/c}$, during which the \Jp particles have a mean free path of zero, in a Hubble-like 
expansion.

\begin{figure}
\includegraphics[width=8cm]{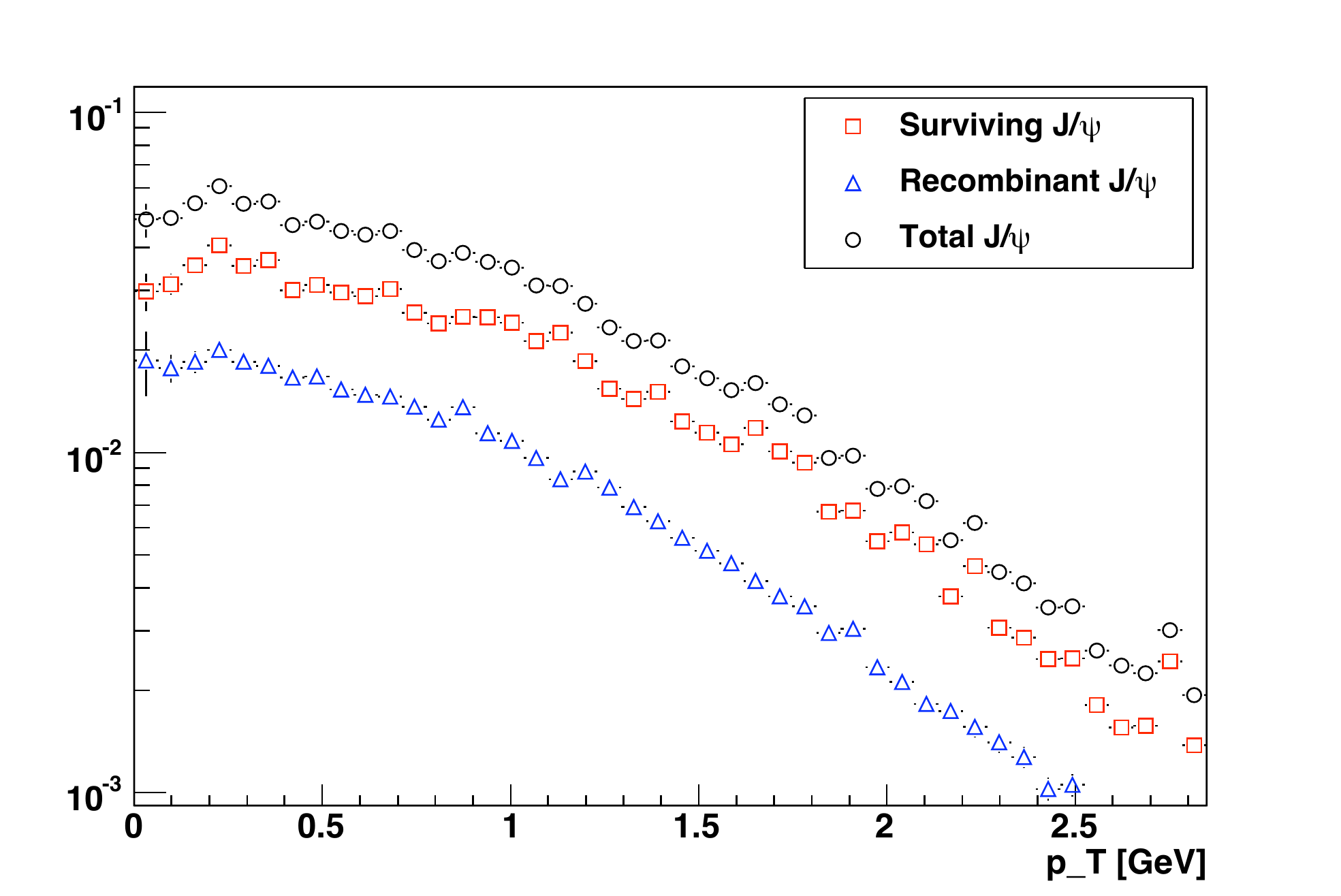}
\caption{ (Color online.)
The surviving and recombinant \Jp yields versus $p_t$.
 }
\label{pt_HF}
\end{figure}

The $p_t$ distributions after this expansion are shown in Figure \ref{pt_HF}. It is visible from this plot 
that the recombinant contribution will observably increase the total yield at low $p_t$ (where the total 
yield is significantly higher than the surviving component alone) and have little effect at higher $p_t$ 
(where the total and the surviving component alone are nearly the same). However, we have found that 
even in this extreme limit, there is no clear signal in the differential $p_t$ yields for there being two 
components for \Jp production at the RHIC.

This test, however, should not be abandoned for measurements of the differential yields at higher 
collision energies. Since the recombinant contribution grows substantially as charm densities are 
increased, it should be checked whether or not the recombinant contribution is more strongly peaked 
in the center of the transverse plane of LHC collisions, and whether or not two components to the 
differential yields should become observable there. We will follow up on this issue in a work we have 
in progress.

\section{Discussion}

\label{conclusions}

We have found that at central Au+Au collisions at  RHIC the fraction of recombinant pairs
should be considerable, up to 30-50\%, with smaller fractions at more peripheral collisions. 
The exact number depends on details of the model, such as duration of the QGP phase and the magnitude of the critical temperature $T_c$.
We have also gone a step further, and attempted to find different behaviors of these two components 
in differential yields, so that these two components might be disentangled. This test (examining the 
differential $p_t$ yields) fails to identify clearly two different components. We will pursue whether or 
not this test works for the yields at the LHC.

Our model for charmonium in sQGP is rather conservative: we merely assume that the constituent 
charm quarks experience dynamics similar to the Langevin dynamics of single charm quarks in SQGP, 
which has already shown good agreement with the $R_{AA}(p_t)$ and $v_2(p_t)$ measured at PHENIX 
for single charm.

One final, careful observation of our results is worth mentioning. As one can see from our results of Fig. \ref{R_AA_rec}, the model seems to be working well for central collisions, in which there is a
QGP phase lasting for several fm/c and leading to a possibility for charm quarks to diffuse away from each other, far enough
so that \Jp states would not survive. However it overpredicts suppression for peripheral collisions, which -- if the cold matter analysis will
hold against further scutiny -- is nearly completely absent. One possible reason for that can be survival of the flux tubes
(QCD strings) between quarks well into the mixed phase or even in small region of temperatures $above$ $T_c$,
as was recently advocated by one of us  \cite{Shuryak:2009cy} in connection with ``ridge" phenomenon.

 \vskip 1.0cm

{\bf Acknowledgments.\,\,}

We thank P. Petreczky for pointing out the issues of setting $T_c$ in our model, which proved to be important in our results.  Our work was partially supported by the US-DOE grants DE-FG02-88ER40388 and
DE-FG03-97ER4014.

\appendix

\appendix

\section{Canonical ensembles for $N$ \barc-pair systems}

\label{canonical}

In this section we will determine a partition function for a canonical ensemble of $N$ charm pair 
systems (that is, an ensemble of very many systems, where each system contains $N$ \barc pairs) 
which correctly averages over different possible pairings of charm and anticharm quarks and can 
therefore describe recombination in heavy-ion collisions. This averaging is possible computationally 
but is non-trivial, and for RHIC collisions we will take a binary approximation which makes this 
averaging much easier. We argue, however, that the unsimplified approach is necessary for describing collisions at the Large Hadron Collider, and for this reason we include this discussion here.

Our simulation could be thought of as a canonical ensemble description of charmonium in plasma: 
we can think of our large set of \barc pairs as a set of systems, each system containing $N$ pairs, 
with each system's dynamics modeled as a stochastic interaction with the medium, with a deterministic 
interaction of each heavy quark with the other quark in the pair. Each system in this set samples the 
distribution of \barc pairs in the initial collision, the geometry of the collision, and also samples the 
stochastic forces on the heavy quarks. Up to this point, we have only thought of each system of this 
set as consisting of a single \barc pair. The interaction of charm quarks from different 
hard events is negligible compared with the stochastic interaction and the interaction within the pair, 
partly because near $T_c$, the dynamics of charm pairs seems best described with some 
generalization of the Lund string model, which allows no interaction between unpaired charm quarks 
\cite{Andersson:1977qx}. Therefore, it is simple bookkeeping to think now of the systems as each consisting of $N$ \barc pairs.

However, even though the dynamics of the system is not changed when considering many \barc pairs 
per collision, the hadronization (``pairing") of these $2N$ charm quarks is now a non-trivial issue. For 
simplicity, assume that the quarks all reach the freezeout temperature $T_c$ at the same proper time. 
There are $N!$ different possible pairings of the quarks and anti-quarks into charmonium states (each 
pairing is an element of the permutation group $S_N$). 
Call a given pairing $\sigma$ (which is an 
element of $S_N$). Near $T_c$, the relative energetics of a pairing $\sigma$ is given by 
\begin{equation}
E_i=\sum_{i}V(|\vec{r}_i-\vec{r}^{'}_{\sigma(i)}|){\rm ,}
\end{equation}
where $V(r)$ is the zero-temperature Cornell potential (with some maximum value at large $r$, 
corresponding to the string splitting), $\vec{r}_i$ the position of the $i$-th charm quark, $\vec{r}^{'}$ 
the position of the $i$-th charm antiquark, and $\sigma(i)$ the integer in the $i$-th term of the 
permutation.

One way to proceed is to average over these pairings according to their Boltzmann factors. In this 
way, the probability of a given pairing would be given by
\begin{equation}
P(i) = \frac{1}{{\cal Z}} \exp(-E_i/T_c){\rm ,} \; {\cal Z}=\sum_{i=1}^{N!} \exp(-E_i/T_c){\rm .}
\end{equation}
However, this averaging ignores the possibility of survival of bound \barc states from the start to 
the finish of the simulation, in that pairings which artificially ``break up" bound states are included 
in the average. This goes against the main point of our last paper: that it is actually the incomplete 
thermalization of \barc pairs which explains the survival of charmonium states.

For this reason, the averaging we perform rejects permutations which break up pairs that would 
otherwise be bound: we average over a subgroup $S_N^{'}$ of $S_N$, and determine the 
probability based on this modified partition function:
\begin{equation}
P(i) = \frac{1}{{\cal Z}} \exp(-E_i/T_c){\rm ,} \; {\cal Z}=\sum_{\sigma \in S_N^{'}} \exp(-E_i/T_c){\rm ,}
\end{equation}
where $E_i$ specifies the energy of a pairing we permit. We will average over the permutations in 
this way. 

By doing this, we will use a fully canonical ensemble description for charm in plasma, which holds 
for any value for $N$, large or small. Previous work in statistical hadronization used the grand 
canonical approach to explain relative abundances of open and hidden charm 
\cite{BraunMunzinger:2000px}, which can only be applied where thermalization may be assumed to be
complete. 

\end{narrowtext}

\end{document}